\documentclass[twocolumn,aps,superscriptaddress,showpacs]{revtex4}
\usepackage{amsthm,amsmath,amssymb}
\usepackage{graphicx}
\usepackage{bm}

\theoremstyle{definition}

\theoremstyle{plain}

\providecommand{\theoremname}{Theorem}
\providecommand{\definitionname}{Definition}
\providecommand{\theoremname}{Theorem}

\begin{document}

\title{Hierarchy of measurement-induced Fisher information for composite
  states}

\author{Xiao-Ming Lu} \email{luxiaoming@gmail.com} \affiliation{Centre for
  Quantum Technologies, National University of Singapore, 3 Science Drive 2,
  Singapore 117543, Singapore}

\author{Shunlong Luo} \email{luosl@amt.ac.cn} \affiliation{Academy of
  Mathematics and Systems Science, Chinese Academy of Science, 100190 Beijing,
  China}

\author{C. H. Oh} \email{phyohch@nus.edu.sg} \affiliation{Centre for Quantum
  Technologies, National University of Singapore, 3 Science Drive 2, Singapore
  117543, Singapore} \affiliation{Department of Physics, National University
  of Singapore, 3 Science Drive 2, Singapore 117543, Singapore}

\begin{abstract}
  Quantum Fisher information, as an intrinsic quantity for quantum states, is
  a central concept in quantum detection and estimation. When quantum measurements are performed on quantum states, classical
  probability distributions arise, which in turn lead to classical Fisher
  information. In this article, we exploit the classical Fisher information
  induced by quantum measurements, and reveal a rich hierarchical structure of
  such measurement-induced Fisher information. We establish a general
  framework for the distribution and transfer of the Fisher information. In
  particular, we illustrate three extremal distribution types of the Fisher
  information: the locally owned type, the locally inaccessible
  type, and the fully shared type. Furthermore, we indicate the significant
  role played by the distribution and flow of the Fisher information in some
  physical problems, e.g., the non-Markovianity of open quantum processes, the
  environment-assisted metrology, the cloning and broadcasting, etc.
\end{abstract}
\pacs{03.65.Ta, 03.67.-a}

\maketitle

\section{Introduction}
Information is always encoded in states, often in the form of
parameters. In order to read out the encoded information, one needs
to distinguish different states, usually via measurements and
parameter estimation. This is the fundamental issue of statistical
distinguishability of states, which exhibits quite different
features for the classical and quantum scenarios~\cite{Fuch}. In the
classical case, the Fisher information is the central notion in
parameter estimation due to the Cram\'er-Rao inequality, which sets
a basic lower bound to the variance of any unbiased estimator in
terms of the Fisher information~\cite{Cra,CF}. Moreover, the maximum
likelihood estimator can saturate the bound asymptotically. In the
quantum realm, the information processing has a much richer
structure and more potentialities which are absent in the classical
world. In particular, in quantum detection and estimation, there are
many (actually infinite) natural quantum extensions of the classical
Fisher information, which are of relevance and significance for
different tasks
~\cite{Woot,Yuan,Helstrom1976,Holevo1982,Petz,Paris}. This is due to
the subtle noncommutative structure of the quantum theory. Among
various versions of quantum Fisher information, two prominent
examples are the one based on the symmetric logarithmic derivatives
~\cite{Helstrom1976,Holevo1982} and the Wigner-Yanase skew
information~\cite{WY,LuoW, LuoH}.

In order to extract information from quantum states, one usually
performs quantum measurements. These quantum measurements induce
naturally classical probability data, and thus lead naturally to the
classical Fisher information, which is deeply connected to the
quantum Fisher information as revealed by the Braunstein-Caves
theorem~\cite{Braunstein1994}. More precisely, for a parameterized
quantum states $\rho _\theta$, a celebrated version of quantum
Fisher information is~\cite{Helstrom1976,Holevo1982}
$$ F (\rho_{\theta}):=  {\rm tr} \rho_{\theta}L_{\theta}^{2} $$
based on the symmetric logarithmic derivative $L_\theta,$ which in turn is
Hermitian and determined by
$\frac{\partial}{\partial\theta}\rho_{\theta}=\frac{1}{2}(L_{\theta}\rho_{\theta}+\rho_{\theta}L_{\theta}).$

If a measurement, described by a positive-operator-valued measure
(POVM) $M =\{ M _{i}| M _{i}\geq0,\sum_{i} M _{i}={\bf 1}\},$ is
performed on the states $\rho _\theta$, then a parameterized
classical probability distribution $p_\theta$ arises with
$$p_\theta (i):={\rm tr}\rho _\theta  M _i.$$
For this family of measurement-induced classical probability distributions, we
have the classical Fisher information
\begin{equation}
  F(\rho_{\theta}|  M ):=\sum_{i}p_{\theta}(i) \left
    (\frac{\partial\ln p_{\theta}(i)}{\partial\theta}\right )^{2}.
  \label{MIF}
\end{equation}
If a composite state $\tilde \rho _\theta$ on the tensor product Hilbert space
$H\otimes K$ is an extension of $\rho_\theta$ in the sense that ${\rm tr}_K
\tilde \rho _\theta =\rho _\theta$, then
\begin{equation}
  F(\tilde \rho _{\theta}|  M\otimes {\bf 1}^K )=F(\rho _{\theta}|  M)
  \label{EMIF},
\end{equation}
where ${\bf 1}^K$ is the identity operator on $K$ and $M\otimes {\bf
  1}^K=\{M_i\otimes {\bf 1}^K\}$ is a POVM on $H \otimes K$.

The elegant Braunstein-Caves theorem states that~\cite{Braunstein1994}
\begin{equation}
  F (\rho _\theta )=\sup _ M F(\rho _\theta| M ), \label{BC}
\end{equation}
where the supremum is over all POVMs. In a word, the quantum Fisher
information, as defined via the symmetric logarithmic derivative, is the
maximum of measurement-induced (classical) Fisher information.

However, in practice, due to various limitations, measurements are
often restricted to certain classes, such as local measurements if
the system is composite. In such cases, the quantum Fisher
information is usually unachievable, and the question arises as how
much Fisher information can be extracted via restricted
measurements. In the multi-partite system scenario, it is also
relevant and important to investigate the distribution and transfer
of the Fisher information over various subsystems. This leads us to
the study of measurement-induced Fisher information. In this
article, we will exploit measurement-induced Fisher information in
order to reveal certain intrinsic structures of the underlying
quantum states, with implications for practical issues.

The article is structured as follows. In Sec.~II, we investigate the
hierarchical structure and fundamental properties of measurement-induced
Fisher information. In Sec.~III, we address the issue concerning distribution
of the Fisher information. In Sec.~IV, we discuss transfer of the Fisher
information in some practical problems. Finally, Sec.~V concludes with a
discussion.

\section{Measurement-induced Fisher Information}

For composite systems, a technical limitation of the implementation of POVMs
may naturally be induced by the tensor product structure of Hilbert spaces,
e.g., for distantly separated subsystems, the local measurements are often easier than the
global measurements. To investigate the distribution of the Fisher information
over composite systems, we first introduce the accessible Fisher information
when the POVMs are restricted. Let ${\cal M}$ be a set of POVMs (implying that
we have technical limitation to implement all the POVMs), we define ${\cal
  M}$-induced Fisher information
\begin{equation}
  F (\rho_{\theta}|{\cal M}):=\sup _{ M  \in {\cal M}}F(\rho_{\theta}| M ),
\end{equation}
where the supremum is over $M =\{ M _{i}| M _{i}\geq0,\sum_{i} M _{i}={\bf
  1}\}$ in ${\cal M},$ and $F(\rho_{\theta}| M )$ is defined via
Eq.~(\ref{MIF}). Obviously, if ${\cal M}$ is the entire set of all POVMs, then
$F(\rho_{\theta}|\mathcal{M})$ equals the quantum Fisher information
$F(\rho_{\theta})$ due to Eq.~(\ref{BC}). Since the classical Fisher
information is convex, it follows that the measurement-induced Fisher
information is convex in the sense that for any set ${\cal M}$ of POVMs,
  $$F(\lambda \rho _\theta +(1-\lambda)\sigma _\theta |{\cal M} )\leq \lambda F(\rho _\theta |{\cal M}) + (1-\lambda) F(\sigma _\theta |{\cal M}).$$
  Here $\lambda \in [0,1]$ is a real number.

  We consider a bipartite system with the Hilbert space $H =H ^{ a }\otimes H
  ^{ b }$ and a parameterized family of bipartite states $\rho _\theta$ with
  reduced states $\rho ^a_\theta :={\rm tr} _b\rho _\theta $ and $\rho
  ^b_\theta :={\rm tr} _a\rho _\theta $. In this context, some natural sets of
  POVMs are listed in Tab.~\ref{tab:notation}, see also
  Ref.~\cite{Hayashi2006}: $\mathcal{M}^a$ and $\mathcal{M}^b$ are sets of all
  local measurements on parties $a$ and $b$, respectively; $\mathcal{M}^{a,b}$
  is the set of all joint POVMs for which $a$ and $b$ perform independently on
  their respective local systems; $\mathcal{M}^{a \rightarrow b }$ is the set
  of joint POVMs that party $b$ performs after party $a,$ conditioned on the
  outcomes of party $a$. Finally, $\mathcal{M} ^{ab}$ is the entire set of all
  POVMs on the composite system.

  \begin{table}
    \caption{\label{tab:notation}Some natural subsets of POVMs for the  composite
      system $H ^{ a } \otimes H ^{ b }.$ }
    \begin{ruledtabular}
      \begin{tabular}{ll}
        POVMs & Set of POVMs  \tabularnewline
        \hline
        $M^a = \{ M _{i}^{a} \otimes {\bf 1} ^{b} \}$ & $ {\cal M} ^{ a}$  \tabularnewline
        $M^b = \{{\bf 1}^{ a }\otimes  M _{i}^{ b }\}$ & $ {\cal M} ^{b}$ \tabularnewline
        $M^{a,b} = \{M _{i}^{a}\otimes  M _{j}^{ b }\}$ & $ {\cal M} ^{a ,b}$ \tabularnewline
        $M^{a \rightarrow b} = \{ M _{i}^{a} \otimes M_{j}^{b|i}\}$ & $ {\cal M} ^{a \rightarrow b}$ \tabularnewline
        $M^{a \leftarrow b} = \{ M _{i}^{a|j} \otimes M_{j}^{b}\}$ & $ {\cal M} ^{a \leftarrow b}$ \tabularnewline
        $M^{a b}=\{M_{i}^{ab}\}$  & ${\cal M} ^{a b}$ \tabularnewline
      \end{tabular}
    \end{ruledtabular}
  \end{table}

  The measurement-induced Fisher information has a natural hierarchical
  structure: If ${\cal M}\subseteq {\cal N},$ then
$$ F(\rho ^\theta |{\cal M})\leq F (\rho ^\theta |{\cal N}).$$
In particular, from the inclusion relations
\begin{equation*}
  \left( {\cal M} ^{a} \cup {\cal M} ^{b}\right)
  \subset {\cal M} ^{a ,b}
  =\left( {\cal M} ^{ a \rightarrow b} \cap {\cal M} ^{ a \leftarrow b } \right)
  \subset {\cal M} ^{a b},
\end{equation*}
we are led to the following hierarchical relations
\begin{eqnarray*}
  F (\rho _\theta |{\cal M}^a) & \leq &  F(\rho _\theta |{\cal M}^{a,b}) \leq F (\rho _\theta |{\cal M}^ {a\rightarrow b})
  \leq F (\rho _\theta |{\cal M}^{ab}),\\
  F (\rho _\theta |{\cal M}^b)  &\leq &   F(\rho _\theta |{\cal M}^{a,b}) \leq F (\rho _\theta |{\cal M}^ {a\leftarrow b})
  \leq F (\rho _\theta |{\cal M}^{ab}).
\end{eqnarray*}
The measurement-induced Fisher information is connected to the corresponding
quantum Fisher information through the Braunstein-Caves theorem and
Eq.~(\ref{EMIF}):
\begin{eqnarray*}
  F (\rho_{\theta}|{\cal M}^{ab}) & = &F (\rho_{\theta}),\label{eq:X_QFI_AB}\\
  F (\rho_{\theta} |{\cal M}^a) & =& F (\rho_{\theta}^{ a }),\label{eq:X_QFI_A}\\
  F (\rho_{\theta}|{\cal M}^b) &
  = &F (\rho_{\theta}^{ b }).\label{eq:X_QFI_B}
\end{eqnarray*}
Meanwhile, the measurement-induced Fisher information for the adaptive
measurements can be expressed as
\begin{align}
  F (\rho_{\theta}| {\cal M}^{a \rightarrow b} ) & =\max_{ M ^{ a }\in {\cal
      M}^a }
  F (\rho_{\theta, M^a} ),\label{eq:X_QFI_A2B}\\
  F (\rho_{\theta} |{\cal M}^ {a\leftarrow b } ) & =\max_{M^b\in {\cal M}^b} F
  (\rho_{\theta, M^b} ),\label{eq:X_QFI_B2A}
\end{align}
where $$\rho_{\theta, M^a}:=\sum_{i}p^a_\theta (i) |i\rangle \langle i|\otimes
\rho ^{b|i}_\theta $$ is the derived classical-quantum state induced by the
local measurement $M ^{a}$, and
$$\rho_{\theta}^{ b |i}:= \frac {{\rm tr_{a}} (M _{i}^{ a }\otimes {\bf
    1}^{ b }) \rho_{\theta}} {p_{\theta}^{ a }(i)}$$ is the conditional state
of party $b$ corresponding to the outcome $i$ after party $a$ performs
measurement $M^a$, while
$$p_{\theta}^{ a }(i):={\rm tr}(M _{i}^{ a }\otimes {\bf 1}^{ b
})\rho_{\theta}$$ is the probability for obtaining outcome $i$, ${\bf 1}^b$ is
the identity operator on $H^b$. The derived state $\rho_{\theta, M^b}$ is
defined similarly.

To establish Eqs.~(\ref{eq:X_QFI_A2B}) and (\ref{eq:X_QFI_B2A}), note that the
joint probability distribution under an adaptive POVM measurement can be
decomposed as
\begin{eqnarray*}
  p_{\theta}(ij)&:=& {\rm tr}( M _{i}^{ a }\otimes  M _{j}^{b|i})\rho_{\theta}\\
  &=&{\rm tr}(M_{i}^{ a }\otimes  {\bf 1}^b)\rho_{\theta} ({\bf 1}^a\otimes M _{j}^{b|i})\\
  &=&{\rm tr}({\rm tr}_a ((M_{i}^{ a }\otimes  {\bf 1}^b)\rho_{\theta} )M _{j}^{b|i})\\
  &=&p_{\theta}^{a}(i)p_{\theta}^{ b }(j|i),
\end{eqnarray*}
where $p_{\theta}^{ b }(j|i):={\rm tr} \rho ^{b|i}_\theta M_{j}^{b|i}$ is the
conditional probability of party $b$ given $a$.

Substituting $p_{\theta}(ij)=p_{\theta}^{ a }(i)p_{\theta}^{ b }(j|i)$ into
the definition of $F(\rho_{\theta}| M ^{ {a\rightarrow b } })$, we have
\begin{eqnarray}
  & & F(\rho_{\theta}|  M  ^{      {a\rightarrow b }     }) \nonumber \\
  &= & \sum _{ij}p_\theta (ij)\left (\frac{\partial\ln p_{\theta} (ij)}{\partial\theta}\right
  )^{2} \nonumber \\
  & =& \sum_{i}p_{\theta}^{ a }(i)\left (\frac{\partial\ln p_{\theta}^{ a }(i)}{\partial\theta}\right )^{2}\nonumber \\
  & &+\sum_{i}p_{\theta}^{ a }(i)\sum_{j}p_{\theta}^{ b }(j|i)\left (\frac{\partial\ln p_{\theta}^{ b }(j|i)}{\partial\theta}\right )^{2}\nonumber \\
  & & +2 \sum_{ij}p_{\theta}^{a}(i)p_\theta ^{b}(j|i) \frac{\partial\ln p_{\theta}^{ a }(i)}{\partial\theta}
  \frac{\partial\ln p_{\theta}^{ b }(j|i)}{\partial\theta } \nonumber \\
  &= &  F(\rho_{\theta}^{ a }|  M  ^{     a      })+\sum_{i}p_{\theta}^{ a }(i)F(\rho_{\theta}^{ b |i}|  M  ^{ b
    |i}) \nonumber \\
  &  & + 2 \sum_{ij} \frac{\partial p_{\theta}^{ a }(i)}{\partial\theta}
  \frac{\partial  p_{\theta}^{ b }(j|i)}{\partial\theta } \nonumber \\
  &=&F(\rho_{\theta}^{ a }|  M  ^{     a      })+\sum_{i}p_{\theta}^{ a }(i)F(\rho_{\theta}^{ b |i}|  M  ^{ b
    |i}).
  \label{eq:QFI_adaptive_1}
\end{eqnarray}
In the last equation, we have used the fact that $\sum _j \frac{\partial
  p_{\theta}^{ b }(j|i)}{\partial\theta }=0$ which is implied by $\sum
_jp^b_\theta (j|i)=1.$

On the other hand, the right hand side of Eq.~(\ref{eq:X_QFI_A2B}) without the
maximization can be calculated by the symmetric logarithmic derivative
\begin{equation*}
  l_{\theta}  =\sum_{i}|i\rangle\langle
  i|\otimes\left(\frac{\partial\ln
      p_{\theta}(i)}{\partial\theta}{\bf 1}^b+L_{\theta}^{ b |i}\right),
\end{equation*}
of $\rho_{\theta,M^a}=\sum_{i}p_{\theta}^{ a }(i)|i\rangle\langle
i|\otimes\rho_{\theta}^{ b |i}$, where $L_{\theta}^{ b |i}$ is the symmetric
logarithmic derivative of $\rho_{\theta}^{b|i}$. Consequently, we
\begin{eqnarray}
  F (\rho_{\theta, M^a} )  & = & {\rm tr} (l_{\theta}^{2} \rho_{\theta, M^a}) \nonumber \\
  &=&F(\rho_{\theta}^{ a }|  M  ^{ a })+\sum_{i}p_{\theta}^{ a }(i) F (\rho_{\theta}^{ b |i}) \label{eq:QFI_adaptive_2}
\end{eqnarray}
Combining Eqs.~(\ref{eq:QFI_adaptive_1}) and (\ref{eq:QFI_adaptive_2}), and
making use of the Braunstein-Caves theorem for every optimization over $ M ^{
  b |i}$, we obtain Eq.~(\ref{eq:X_QFI_A2B}). Eq.~(\ref{eq:X_QFI_B2A}) can be
obtained similarly.

Measurement-induced Fisher information for any set ${\cal M}$ of POVMs, listed
in Tab. \ref{tab:notation}, is monotonic under any local quantum operations
$E^a$ and $E^b$ on $a$ and $b,$ respectively, in the sense that
\begin{equation}
  F (E ^a \otimes E^b (\rho_{\theta}) |{\cal M}) \leq F (\rho_{\theta} |{\cal M}).\label{eq:monotonicity_X_QFI}
\end{equation}

To establish this, note that the classical Fisher information
$F\left(\rho_{\theta}| M \right)$ is a functional of the probability
distribution $p_{\theta}(i)= {\rm tr} \rho_{\theta} M _{i}.$ For any quantum
operation $E (X)=\sum_{\mu}E_{\mu}X E_{\mu}^{\dagger}$ with $E_{\mu}$ the
Kraus operators satisfying $\sum_{\mu} E_{\mu}^{\dagger} E_{\mu}={\bf 1},$ we
have $ {\rm tr}E (\rho_{\theta}) M _{i}= {\rm tr}\rho_{\theta} E ^{\dagger}( M
_{i})$, where $E^{\dagger}(X)=\sum_{\mu} E_{\mu}^{\dagger}X E_{\mu}$ is the
adjoint operation of $E$. Furthermore,
\begin{equation*}
  F (E(\rho_{\theta}) | M )=F ( \rho_{\theta}| E^ {\dagger}(M)),
\end{equation*}
where $ E^{\dagger}(M):=\{E^{\dagger}( M _{i})\}$ is also a POVM. Therefore,
for any set ${\cal M}$ of operations,
\begin{eqnarray*}
  F (E (\rho_{\theta})| {\cal M} ) & := & \max_{  M  \in {\cal M} }F ( E (\rho_{\theta})|  M  )\\
  & = & \max_{  M  \in {\cal M} } F ( \rho_{\theta} | E^{\dagger}(M) ) \\
  & = & \max_{  M  \in E ^{\dagger}( {\cal M})}F ( \rho_{\theta}|  M ),
\end{eqnarray*}
where $E ^{\dagger}({\cal M}):=\{E ^{\dagger}( M )| M \in {\cal M}\}$. In
particular, for any set ${\cal M}$ of POVMs listed in Table I, one easily
checks that for $E=E^a\otimes E^b$,
$$E^\dagger({\cal M}) \subseteq {\cal M}, $$
from which the desired inequality~(\ref{eq:monotonicity_X_QFI}) follows.

\section{Distribution of Fisher Information \label{sec:distr-fish-inform}
} In this section, we consider the distribution of
measurement-induced Fisher information within bipartite systems and
illustrate some extremal cases.

First, consider the relation between the locally accessible Fisher
information $$ F(\rho_{\theta}|{\cal M}^a )= F (\rho_{\theta}^{ a
}), \qquad F (\rho_{\theta}|{\cal M}^b)= F (\rho_{\theta}^{ b }),$$
which are also called marginal Fisher information hereafter, and the
global quantum Fisher information $$ F(\rho_{\theta}|{\cal M}^{ab})=
F (\rho_{\theta}).$$ For any product state
$\rho_{\theta}=\rho_{\theta}^{ a }\otimes\rho_{\theta}^{ b },$ the
quantum Fisher information is additive in the sense that $F
(\rho_{\theta})= F (\rho_{\theta}^{ a })+ F (\rho_{\theta}^{ b })$.
For such cases, we say that the Fisher information of bipartite
states $\rho_\theta$ is (exclusively) locally owned. However, for
general bipartite states, due to correlations between the two
parties $a$ and $b$, the quantum Fisher information can be either
superadditive, i.e., $ F (\rho_{\theta})\geq F (\rho_{\theta}^{ a
})+ F (\rho_{\theta}^{ b })$, or subadditive, i.e., $ F
(\rho_{\theta})\leq F (\rho_{\theta}^{ a })+ F (\rho_{\theta}^{ b
})$. Roughly speaking, the Fisher information may be distributed
exclusively in local parties, shared between the two parties,
inaccessible to local parties, or in a more complex manner. The
mechanism about how correlations affect the distribution of the
quantum information seems unclear. In the following, we elucidate
two extremal scenarios for the distribution of Fisher information:
locally inaccessible and fully shared.

We say that Fisher information of bipartite states $\rho _\theta$ is locally
inaccessible if $ F (\rho_{\theta}^{ a })= F (\rho_{\theta}^{ b })=0$ while
$F(\rho_{\theta})\neq 0.$ A simple example of this type distribution of Fisher
information is $\rho_{\theta}=|\Psi_{\theta}\rangle\langle\Psi_{\theta}|$ with
\begin{equation*}
  |\Psi_{\theta}\rangle=\frac 1{\sqrt 2}(|0^{ a }\rangle\otimes|0^{ b }\rangle+e^{i\theta}|1^{ a } \rangle \otimes|1^{ b }\rangle ),
\end{equation*}
Here, it can be easily checked that $$F(\rho_{\theta})=1/2, \quad F
(\rho_{\theta}^{ a })= F (\rho_{\theta}^{ b })=0,$$ and for such states, the
Fisher information is superadditive in the sense that
 $$ F (\rho_{\theta})> F (\rho_{\theta}^{ a })+ F (\rho_{\theta}^{ b }).$$

 We say that the Fisher information of $\rho_\theta $ is fully sharable if $ F
 (\rho_{\theta})= F (\rho_{\theta}^{ a })= F (\rho_{\theta}^{ b }) \ne 0.$ Two
 simple examples of such type distribution of Fisher information are as
 follows. The first is the classically correlated bipartite states
 \begin{equation*}
   \rho_{\theta}=\sum_{i}p_{\theta}(i)|i^{ a }\rangle\langle i^{ a }|\otimes|i^{ b }\rangle\langle i^{ b }|.
 \end{equation*}
 Here, it can be readily checked that
 \begin{equation*}
   F (\rho_{\theta})= F (\rho_{\theta}^{ a })= F (\rho_{\theta}^{ b })=\sum_{i}p_{\theta}(i)
   \left (\frac{\partial\ln p_{\theta}(i)}{\partial\theta}\right )^{2},
 \end{equation*}
 which is exactly the classical Fisher information of the probability
 distribution $\{p_{\theta}(i)\}.$

 The second example of the fully shared type is
 \begin{equation*}
   \Psi_{\theta}=\cos\frac{\theta}{2}|0^{ a }\rangle\otimes|0^{ b }\rangle+\sin\frac{\theta}{2}|1^{ a }\rangle\otimes|1^{ b }\rangle,
 \end{equation*}
 for which one easily gets $$ F (\rho_{\theta})= F (\rho_{\theta}^{ a })= F
 (\rho_{\theta}^{ b })=1.$$ It is evident that for the above two examples, the
 Fisher information is subadditive in the sense that
 $$ F (\rho_{\theta}) < F (\rho_{\theta}^{ a })+ F (\rho_{\theta}^{ b }).$$

 \section{Transfer of Fisher Information}

 Based on the distribution of measurement-induced Fisher information, one may
 further investigate the flow of the Fisher information over composite
 systems. Many of the existing physical problems can be elucidated through the
 flow of the Fisher information , e.g., the non-Markovianity of open quantum
 processes~\cite{Lu2010}, the environment-assisted
 precision measurement~\cite{Goldstein2011,Cappellaro2012}, and the
 cloning/broadcasting of the Fisher information~\cite{Lu2012}, etc. It will be
 interesting to investigate how measurement-induced Fisher information is
 transferred in such practical issues. Due to the rich hierarchy of the
 measurement-induced Fisher information, in general, the Fisher information
 will not only flow between the reduced systems, but also between the shared type
 and locally inaccessible type. In this section, we illustrate various features of
 transfer of the Fisher information though several typical examples.

 {\sl Example} 1. First, let us consider a scenario in which the Fisher
 information is transferred from the (exclusively) locally owned type into the
 shared type. We assume that the initial Fisher information about the
 parameter $\theta$ is only locally owned by party $a$, and the parameterized
 family of bipartite states is assumed as product states $\sigma^a_\theta
 \otimes \sigma^b$. After a unitary transformation $U$ on the bipartite
 system, the output states read
 $$ \rho_\theta = U(\sigma^a_\theta\otimes\sigma^b) U^\dagger. $$
 The global Fisher information remains unchanged under the unitary transformation. If we
 further assume the spectral decomposition form $$\sigma^a_\theta = \sum_i p_\theta(i)|i^a\rangle\langle
 i^a|,$$ and that $U$ has a conditional form as $$U=\sum_{i}|i^a\rangle \langle
 i^a|\otimes U^b_i,$$  then the two reduced states of the composite system read as
 $$
 \rho^a_\theta=\sigma^a_\theta \quad \mbox{and}\quad \rho^b_\theta=\sum_i
 p_\theta(i)U^b_i\sigma^b U_i^{b\dagger},
 $$
 from which we see that the reduced states of party $a$ remain unchanged, and thus the marginal Fisher formation for party $a$ remains the same.
 However, party $b$ gains some Fisher information, because its reduced
 state become dependent on the parameter $\theta$. This indicates
 the transfer of some Fisher information from the locally owned
 type to the shared type. An extremal situation for this kind of Fisher information
 transfer is the broadcasting of the quantum Fisher information, where all of
 the locally owned type of Fisher information is transferred to
 the shared type, see Ref.~\cite{Lu2012}.

{\sl Example} 2. Now we consider a scenario in which  the Fisher
 information transfers from the locally owned type to the locally
 inaccessible type. We choose the input states as
 $\sigma_\theta=|\Omega_\theta\rangle\langle\Omega _\theta |$, where
$$|\Omega _\theta\rangle = \frac{1}{\sqrt{2}}\left(|0^a\rangle+e^{i\theta}|1^a\rangle\right) \otimes
|0^b\rangle.$$ After the controlled-NOT operation
\begin{equation}
  \label{eq:CNOT}
  U=|0^a\rangle\langle
  0^a|\otimes \mathbf{1}^b+|1^a\rangle\langle 1^a|\otimes
  \left(|0^b\rangle \langle 1^b|+|1^b\rangle \langle 0^b|\right),
\end{equation}
the output states read $\rho_\theta :=U\sigma _\theta U^\dagger
=|\Psi_\theta\rangle\langle \Psi_\theta|$ with
$$
|\Psi_\theta\rangle = \frac{1}{\sqrt{2}}\left(|0^a\rangle \otimes
|0^b\rangle+e^{i\theta}|1^a\rangle \otimes | 1^b\rangle \right).
$$
For the input state, we have $F(\sigma^a_\theta)=F(\sigma_\theta)=1$
and $F(\sigma^b_\theta)=0$, which means that the Fisher information
is (exclusively) locally owned by party $a$. For the output state,
we have $F(\rho_\theta)=1$ and
$F(\rho^a_\theta)=F(\rho^b_\theta)=0$, and thus the Fisher
information is locally inaccessible to both parties $a$ and $b$.

{\sl Example} 3. Finally, we consider a situation for the transfer
of the Fisher information from the locally inaccessible type to the
locally accessible type. This kind of Fisher information transfer is
strongly related to some practical proposals on enhancing the
sensitivity of precision measurement with the help of  ancillary
systems \cite{Goldstein2011,Cappellaro2012}. As a simple
illustration, consider the following input states
$\sigma_\theta=|\Omega _\theta\rangle\langle \Omega _\theta|$ with
$$|\Omega _\theta\rangle
=\frac{1}{\sqrt{2}}\left(|0^a\rangle \otimes
|0^b\rangle+e^{i\theta}|1^a\rangle \otimes |1^b\rangle\right).$$
Such states often emerge in the field of quantum metrology as the
optimal states for high-precision measurements
\cite{Giovannetti2006}. After the controlled-NOT operation $U$ given
by Eq.~(\ref{eq:CNOT}), the output states read $\rho_\theta:=U\sigma
_\theta U^\dagger =|\Psi_\theta\rangle \langle \Psi_\theta|$ with
$$|\Psi_\theta\rangle =\frac{1}{\sqrt{2}}\left(|0^a\rangle +
  e^{i\theta}|1^a\rangle \right)\otimes |0^b\rangle.$$ Actually, this is the
reversal of the process in Example 2. One can easily check that for
the input states the Fisher information is inaccessible to both
parties $a$ and $b$, meanwhile in the output states the Fisher
information becomes accessible to party $a$. This kind of Fisher
information transfer may be called the concentration of Fisher
information, compared with the broadcasting of Fisher information.
The concentration of Fisher information from the composite system
into the measurable subsystems may help to improve measurement
precision.

\section{Discussion}
Though quantum Fisher information is intrinsic to quantum states, it
may not be fully extractable if quantum measurements are restricted,
which often occurs in practice. Thus it is of basic importance to
study how much Fisher information can be derived via certain class
of quantum measurements. This issue concerning measurement-induced
Fisher information is briefly investigated here. Based on the
distribution and hierarchical structure of measurement-induced
Fisher information,  the flow of Fisher information over composite
systems is further investigated. Many physical problems can be
elucidated through the flow of Fisher information, such as the
non-Markovianity of open quantum dynamics, the environment-assisted
precision measurement, and the cloning and broadcasting of parameter
information via quantum channels.

We point out that measurement-induced Fisher information provides a
convenient and meaningful tool for quantifying correlations and
nonlocality, and that an initial approach for using the
Wigner-Yanase skew information (a version of quantum Fisher
information) in quantifying correlations is already put forward
in~\cite{LFOh}. The hierarchical structure of measurement-induced
Fisher information sheds alternative lights into correlations and may
be exploited for the purpose of revealing correlations from the
perspective different from the conventional entropic approach.

We have only considered a single parameter case. From both theoretical and
practical points of view, it will also be interesting and useful to study the
multi-parameter case, which will certainly exhibit a more complex structure
for measurement-induced Fisher information.

\begin{acknowledgments}
  X.-M. Lu thanks X. Wang for helpful discussions. This work was supported
  by National Research Foundation and Ministry of Education, Singapore, Grant
  No. WBS: R-710-000-008-271, and by the Science Fund for Creative Research
Groups, National Natural Science Foundation of China, Grant No.
10721101, the National Center for Mathematics and Interdisciplinary
Sciences, Chinese Academy of Sciences, Grant No. Y029152K51.
\end{acknowledgments}

\end{document}